\documentclass{svproc}
\usepackage{amsmath}
\usepackage{hyperref}
\usepackage{amssymb}
\usepackage[force]{feynmp-auto}
\usepackage{graphicx}
\usepackage{subcaption}

\newcommand*\he{$^3\mathrm{He}$}
\newcommand*\V[1]{\mathbf{#1}}
\newcommand*\eps{\epsilon}
\newcommand*\sml{\mathsf{s}}
\newcommand*\kb{k_{\mathrm{B}}}
\newcommand*\cO{\mathcal{O}}
\renewcommand{\Re}{\mathfrak{Re}}

\newcommand*\vv{\V{v}}
\newcommand*\vp{\V{p}}
\newcommand*\vR{\V{R}}
\newcommand*{\atanfactor}{\arctan\frac{x_{\mathrm{c}}}{\sqrt\vartheta}}
\begin{document}
\mainmatter
\title{Pairing Fluctuation Corrections to the Kinetic Theory of Liquid $^3\mathrm{He}$}
\titlerunning{Fluctuation Corrections to Zero Sound in $^3\mathrm{He}$}
\author{Wei-Ting Lin and J. A. Sauls$^*$}
\institute{
Department of Physics and Astronomy \\
Center for Applied Physics \& Superconducting Technologies\\
Northwestern University, Evanston, IL 60208, USA}
\maketitle
\begin{abstract}
Liquid \he{} is a Fermi liquid that undergoes a BCS-type phase transition to a spin-triplet superfluid, making it valuable for understanding interacting fermions. When the temperature approaches the transition temperature $T_{\mathrm{c}}$ from above, physical properties can be modified by Cooper pair fluctuations, leading to deviations from the predictions of Fermi liquid theory. We use nonequilibrium Green's function theory to study the role of pair fluctuations on quasiparticle transport. The Boltzmann-Landau kinetic equation, which describes the transport properties of Fermi liquids, acquires corrections from the interaction of quasiparticles with pair fluctuations. As an application, we study the effects of pair fluctuations on the propagation of zero sound in liquid \he{}. We calculate the temperature-dependent correction to the zero sound velocity due to pair fluctuations, and compare the result with existing experimental data.
\end{abstract}
%
\section{Introduction}
In condensed matter physics \he{} is a special system. Liquid $^3$He becomes a quantum liquid below $\sim 1\,\mathrm{K}$, exhibiting the thermodynamic and transport properties of a Fermi liquid predicted by Landau~\cite{lan56,lan57,dobbs00}. Liquid \he{} has inspired studies on, and provided insight into, other condensed matter systems, such as electrons in strongly interacting metals. For example, the quasiparticle-quasiparticle scattering rate in a Fermi liquid is predicted to be proportional to $T^2$, which has been observed in liquid \he{} by measuring the attenuation of zero sound~\cite{abe66}. Similarly, $T^2$ behavior of the resistivity is observed in high-purity heavy-electron metals~\cite{and75,kyc98}.
Liquid \he{} undergoes a superfluid phase transition at temperatures below $T_{\mathrm{c}}\simeq 2.5\,\mathrm{mK}$~\cite{osh72}, driven by a mechanism similar to BCS pairing in superconductors (Boson-mediated attraction), but in the case of liquid \he{} Cooper pairs form in a superposition of spin-triplet, p-wave states~\cite{leg75}. Indeed liquid \he{} was the first realization of unconventional superconductivity in which the pairing state breaks spin- and space-time symmetries in addition to global $\mathrm{U}(1)$ symmetry~\cite{vollhardt90}.

Although both the Fermi liquid and superfluid phases of \he{} have been investigated extensively~\cite{dobbs00}, the \emph{border} between these phases is not so well-studied. By contrast, for superconductors, the resistivity in the normal phase can decrease below the normal value when the temperature is close to the transition temperature. This phenomenon, called \emph{paraconductivity}, is due to the enhanced conductivity of long-lived Cooper pair fluctuations near the superconducting phase transition, and has been studied extensively~\cite{sko75}. The fact that Landau's Fermi liquid theory and the BCS pairing theory underlie the physical properties of both liquid \he{} and that of electrons in metals suggests that pairing fluctuations should also influence the physical properties of liquid \he{} near the superfluid transition temperature, $T_{\mathrm{c}}$. 

Shortly after the discovery of the superfluid phases of liquid \he{} Emery proposed that measurements of the effects of pairing fluctuations on transport properties might be used to determine the orbital angular momentum of the Cooper pairs~\cite{eme76}.
Emery's proposal was based on calculations of the corrections to transport properties, e.g. the viscosity of normal \he{}, based on a heuristic proposal for the collision integral of the Boltzmann-Landau transport equation.
Parpia et~al.~\cite{par78} did observe a deviation of the viscosity compared to the prediction of Fermi liquid theory, but only within a tiny window of temperatures, $\sim 10^{-3}T_{\mathrm{c}}$, near the transition, whereas Emery predicted a wider temperature range for the fluctuation corrections. 
About the same time Paulson and Wheatley~\cite{pau78c} observed enhanced attenuation of zero sound over a temperature window of order $\sim 10^{-1}T_{\mathrm{c}}$, and shortly thereafter Samalam and Serene~\cite{sam78} used Emery's collision integral to calculate the pair fluctuation correction the zero sound attenuation and obtained a result consistent with Paulson and Wheatley's data.

One motivation for this work was to formulate a theory of quasiparticle-pair-fluctuation scattering for the nonequilibrium properties of a Fermi liquid near a superconducting/superfluid transition from a first-principles framework of quantum transport theory. A second motivation was the experimental observation of changes in the velocity of zero sound for temperatures near the superfluid transition in \he{}~\cite{lee96}. Pair-fluctuation corrections to the \emph{velocity} of zero sound are not accounted for within the formulation of Emery, Samalam and Serene, as shown by Pal and Bhattacharyya~\cite{pal79}.

In this article we develop a kinetic theory for quasiparticle transport, including the leading order corrections from incipient pair fluctuations above $T_{\mathrm{c}}$. The details of the derivation starting from Keldysh field theory for interacting Fermi systems is outlined in a separate report~\cite{lin21}. A summary of the main result is presented in Sect.~\ref{sec-kinetic}. In Sect.~\ref{sec-zs} we discuss zero sound, a coherent particle-hole excitation of the Fermi surface, and calculate the correction to the zero sound velocity from the interaction of quasiparticles with long-lived pair fluctuations. The collision integral and the zero sound attenuation are discuss in Ref.~\cite{lin21}.

\section{Kinetic Equation with Pair Fluctuation Corrections}\label{sec-kinetic}

\begin{figure}[t]
\begin{minipage}{\textwidth}
\centering
\vspace{2em}
\begin{fmffile}{fmf-Self_Energy_High}
  \begin{fmfgraph*}(100,45)
    \fmfpen{thin}
    \fmfset{arrow_len}{3mm}
    \fmftop{b1}
    \fmfv{label=(a) $\sml^0\hspace*{25mm}$}{b1}
    \fmfleft{l1,l2}
    \fmfright{r1,r2}
    \fmfv{label=$a$,l.d=0.5mm,l.a=180}{l1}
    \fmfv{label=$b$,l.d=0.5mm,l.a=0}{r1}
    \fmfv{decor.shape=circle,d.f=0,d.si=8mm,label=$\Sigma_{p}^{\mbox{\tiny high}}$,l.d=-1.5mm}{v1}
    \fmf{fermion,tension=.9,label=$p$,l.a=+90}{r1,v1}
    \fmf{fermion,tension=.9,label=$p$,l.a=-90}{v1,l1}
  \end{fmfgraph*}
\end{fmffile}
\begin{fmffile}{fmf-Self_Energy_Landau}
  \begin{fmfgraph*}(100,45)
    \fmfpen{thin}
    \fmfset{arrow_len}{3mm}
    \fmftop{b1}
    \fmfv{label=(b) $\sml^1\hspace*{25mm}$}{b1}
    \fmfleft{l1,l2}
    \fmfright{r1,r2}
    \fmfpolyn{empty,tension=.2,label=$\Gamma^{\mathsf{ph}}$}{G}{4}
    \fmf{fermion,tension=.9,label=$p$,l.a=+90}{r1,G1}
    \fmf{fermion,right,tension=0.1,label=$p'$}{G2,G3}
    \fmf{fermion,tension=.9,label=$p$,l.a=-90}{G4,l1}
    \fmfv{label=$a$,l.a=180,l.d=0.5mm}{l1}
    \fmfv{label=$b$,l.d=0.5mm,l.a=0}{r1}
  \end{fmfgraph*}
\end{fmffile}
\begin{fmffile}{fmf-Self_Energy_Binary}
  \begin{fmfgraph*}(120,45)
    \fmfpen{thin}
    \fmfset{arrow_len}{3mm}
    \fmftop{b1}
    \fmfv{label=(c) $\sml^2\hspace*{25mm}$}{b1}
    \fmfleft{l1,l2}
    \fmfright{r1,r2}
    \fmfv{label=$a$,l.d=0.5mm,l.a=180}{l1}
    \fmfv{label=$b$,l.d=0.5mm,l.a=0}{r1}
    \fmfpolyn{empty,tension=0.2,label=$\Gamma$}{G}{4}
    \fmfpolyn{empty,tension=0.2,label=$\Gamma$}{H}{4}
    \fmf{fermion,tension=0.9,label=$p$,l.a=90}{r1,H1}
    \fmf{fermion,tension=0.9,label=$p'''$,l.d=-4mm}{H4,G1}
    \fmf{fermion,tension=0.0,label=$p''$,l.a=90}{G2,H3}
    \fmf{fermion,right,tension=0.0,label=$p'$,l.a=90}{H2,G3}
    \fmf{fermion,tension=0.9,label=$p$,l.a=90}{G4,l1}
  \end{fmfgraph*}
\end{fmffile}
\\
\vspace{4em}
\begin{fmffile}{fmf-Self_Energy_Cooper}
  \begin{fmfgraph*}(100,45)
    \fmfpen{thin}
    \fmfset{arrow_len}{3mm}
    \fmftop{b1}
    \fmfv{label=(d) $\sml^3\hspace*{25mm}$}{b1}
    \fmfleft{l1,l2}
    \fmfright{r1,r2}
    \fmfv{label=$a$,l.d=0.5mm,l.a=180}{l1}
    \fmfv{label=$b$,l.d=0.5mm,l.a=0}{r1}
    \fmfpolyn{empty,tension=.2,label=$\Gamma_{Q}$}{G}{4}
    \fmf{fermion,tension=.9,label=$p$,l.a=90}{r1,G1}
    \fmf{fermion,left,tension=0.1,label=$-p+Q$}{G3,G2}
    \fmf{fermion,tension=.9,label=$p$,l.a=90}{G4,l1}
  \end{fmfgraph*}
\end{fmffile}
\caption{Self-energy diagrams and their order of magnitude in the small expansion 
         parameter, $\sml\in\{\kb T_{\mathrm{c}}/E_{\mathrm{f}},1/k_{\mathrm{f}}\xi\}$ of Fermi liquid theory.
(a) $\Sigma_{p}^{\mbox{\tiny high}}$ is $\cO(\sml^0)$ and gives quasiparticle the mass renormalization,
(b) is the Landau mean field self-energy of $\cO(\sml^1)$,
(c) is the self-energy from binary collision scattering and is $\cO(\sml^2)$, and
(d) is the self-energy for quasiparticles-pair fluctuation interactions,
    is $\cO(\sml^3)$; vertex $\Gamma_Q$ diverges for 
       $T\rightarrow T_{\mathrm{c}}^+$ and $Q\rightarrow 0$.
}
\label{fmf-Self_Energies}
\end{minipage}
\end{figure}
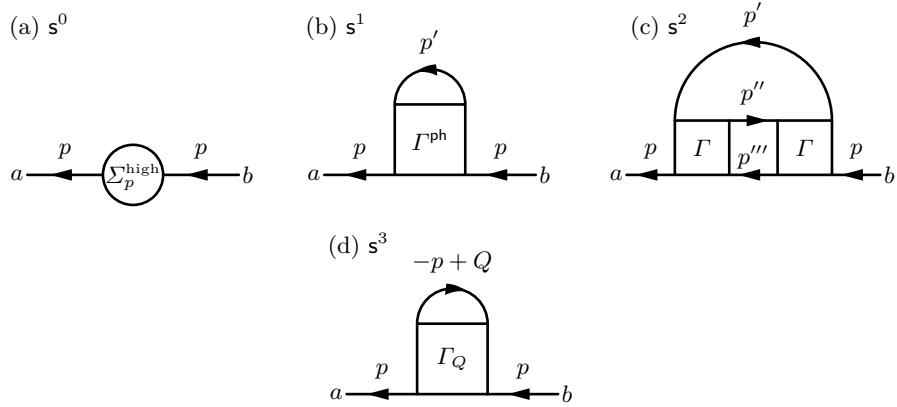

Landau's Fermi liquid theory implicitly assumes that the Fermi sea - the quasiparticle vacuum - is stable, i.e. there is no symmetry-breaking phase transition. Thus, to incorporate the superfluid transition, we need a microscopic theory of the interacting Fermi system. 
The quasiparticle properties, including the distribution function, $n_{\vp}(\vR,t)$, of the interacting Fermi system are obtained from the nonequilibrium single-particle Keldysh Green's function. From Dyson's equation and the quasiparticle \emph{ansatz} for the Fermionic spectral function~\cite{AGD}, it can be shown that the quasiparticle distribution function satisfies~\cite{lin21},
\begin{multline}\label{KE-0}
\left(\partial_t + \frac{\vp}{m}\cdot\nabla_\V{R}\right)n_{\vp}
- \nabla_\V{R}\Re\Sigma^{11}\big|_{\eps=\eps_\V{p}}\cdot\nabla_\V{p}n_{\vp}
+ \nabla_\V{p}\Re\Sigma^{11}\big|_{\eps=\eps_\V{p}}\cdot\nabla_\V{R}n_{\vp} 
\\
- \partial_\eps \Re\Sigma^{11}\big|_{\eps=\eps_\V{p}}\partial_t n_{\vp} 
= 
-i\Sigma^{12}\big|_{\eps=\eps_{\V{p}}}(1-n_{\vp})-i\Sigma^{21}\big|_{\eps=\eps_{\V{p}}} n_{\vp}
\,,
\end{multline}
where $n_{\vp}(\vR,t)n=-i\int d\eps G_p^{12}(\vR,t)$ is a quasiclassical distribution function in phase space $(\vp,\vR)$ for nonequilibrium disturbances that are long-wavelength compared to the Fermi wavelength $\hbar/p_{\mathrm{f}}$ and low-frequency compared to the atomic scale set by the Fermi energy, $E_{\mathrm{f}}/\hbar$.
The terms in the first set of parentheses on the left-hand side of Eq.~\eqref{KE-0} describe the smooth evolution in phase space for noninteracting fermions. Interactions enter the kinetic equation via the Keldysh components of the self-energy term, $\Sigma^{12}$, $\Sigma^{21}$, and $\Sigma^{11}$. The leading order processes contributing to the self energy are represented by the Feynman diagrams shown in Fig.~\ref{fmf-Self_Energies}. 

We include the leading order contributions to the self-energy that typically define an interacting Fermi liquid in Fig.~\ref{fmf-Self_Energies}(a,b,c). These diagrams are derived using the quasiclassical method, and their magnitude can be formally classified by a small parameter, $\sml\in\{\kb T/E_{\mathrm{f}},\hbar/\tau E_{\mathrm{f}},\hbar/p_{\mathrm{f}} l,\hbar\omega/E_{\mathrm{f}}\,\ldots\}$~\cite{ser83,rai94b}. Diagram (a) in Fig.~\ref{fmf-Self_Energies} generates the renormalized Fermi velocity, or effective mass $m^*=p_{\mathrm{f}}/v_{\mathrm{f}}$, while diagram (b) reduces to the Landau mean-field interaction energy, characterized by the Landau parameters $F^{s,a}_l$. These two diagrams are the leading-order contributions to the Fermi liquid theory. Including these two terms in the self-energy in Eq.~\eqref{KE-0}, we obtain the \emph{collisionless} Boltzmann-Landau equation, from which zero sound is a Bosonic eigenmode. The third diagram (c) contributes to the collision integral for binary collisions of quasiparticles defined by a four-point vertex for the scattering amplitude of quasiparticles with momenta and energies evaluated on the Fermi surface~\cite{lin21}. The Keldysh components $\Sigma^{12}$ and $\Sigma^{21}$, which are moved to the right-hand side of Eq.~\eqref{KE-0} contribute the ``scattering out'' and ``scattering in'' terms of the collision integral. Note that there is no contribution to $\Re\Sigma^{11}$ from diagram (c)~\cite{ser83}.

Our focus is on diagram (d) for which the vertex $\Gamma_Q$ represents the sum of ladder diagrams in the particle-particle channel, and is related to the Cooper instability. In particular, $\Gamma_Q$ diverges for total momentum and energy of a pair $Q=0$ and $T\rightarrow T_{\mathrm{c}}^+$ (see the Appendix). Thus, diagram (d), representing the interaction between quasiparticles and pair fluctuations which is formally of order $\sml^3$, becomes significant in the near vicinity of the superfluid transition because of the divergence of the pair amplitude. 

Here we report on the effects of virtual emission and absorption of Cooper pairs on the velocity of zero sound implied by diagram (d). This diagram also contains Keldysh components $\Sigma^{12}$ and $\Sigma^{21}$, and thus contributes to the collision integral and the attenuation of zero sound by the scattering of the coherent particle-hole excitation of zero sound by pair fluctuations.

Combining the self energies represented by diagrams Fig.~\ref{fmf-Self_Energies}(a),(b) and (d) 
with Eq.~\eqref{KE-0} we obtain for the linearized, collisionless Boltzmann-Landau equation,
\begin{multline}
\label{linear-BLE} 
\big(1-\partial_\eps\mathfrak{Re}\Sigma_{\mathrm{fluc}}^{11}\big|_{\eps=\eps_\V{p}}^{\mathrm{eq}}\big)  \partial_t\delta n_{\vp}
+\big(\V{v}_{\V{p}} + \nabla_\V{p}\mathfrak{Re}\Sigma_{\mathrm{fluc}}^{11}\big|_{\eps=\eps_\V{p}}^{\mathrm{eq}}\big)\cdot \nabla_\V{R}\delta n_{\vp} 
\\
-\nabla_\V{R}\big(\delta\eps_{\V{p}}
+\mathfrak{Re}\delta\Sigma_{\mathrm{fluc}}^{11}\big|_{\eps=\eps_\V{p}}\big)\cdot\nabla_\V{p}n_0(\eps_{\vp})
=0\,,
\end{multline}
where $\delta n_{\vp}$ is the deviation from the equilibrium distribution, $n_0(\eps_{\vp})$, $\vv_{\vp}=v_f\hat\vp$ is the Fermi velocity with $v_f=p_f/m^*$, and
\begin{equation}
\delta\eps_{\V{p}} = 
\frac{1}{V} \sum_{\V{p}'\sigma'} f_{\V{p}\sigma, \V{p}'\sigma'} \delta n_{\V{p}'\sigma'}
\end{equation}
is the mean field correction to the quasiparticle energy, where $f_{\vp\sigma,\vp'\sigma'}$ is the Landau quasiparticle interaction. 
The term $\Sigma_{\mathrm{fluc}}\big|^{\mathrm{eq}}$ represents the self-energy contribution from diagram Fig.~\ref{fmf-Self_Energies}(d), evaluated in equilibrium, while $\delta\Sigma_{\mathrm{fluc}}$ is the nonequilibrium correction resulting from $\delta n_{\vp}$.
For temperatures near $T_{\mathrm{c}}$ the influence of the pair-fluctuations leads to corrections to the Fermi velocity, Landau quasiparticle interaction energy, and the zero sound dispersion relation.
In addition, the coefficient of the time derivative in the kinetic equation is modified. The reason for this is that the vertex $\Gamma_Q$ that determines the fluctuation self-energy is sensitive to the low-energy scale. This is in contrast to the leading-order self-energy, where all vertices are insensitive to low-energy variables~\cite{ser83}.
We have omitted the collision integral on the right-hand side of Eq.~\eqref{KE-0}. These terms are discussed in Ref.~\cite{lin21} and give rise pair-fluctuation corrections to zero sound attenuation.

In what follows we report our results for the pair-fluctuation corrections to the Fermi velocity, Fermi liquid parameters and the zero sound velocity; the details of the analysis is reported in Ref.~\cite{lin21}.
The reduced temperature is denoted by $\vartheta\equiv(T-T_{\mathrm{c}})/T_{\mathrm{c}}$, and the Cooper pair coherence length is defined as $\xi_0^2=\frac{7\zeta(3)}{48\pi^2}\frac{v_\mathrm{f}^2}{T_\mathrm{c}^2}$. Note that $(N(0)T_{\mathrm{c}}\xi_0^3)^{-1}$ is the small prefactor determining the magnitude of the pair-fluctuation corrections, where $N(0)$ is the density of state at the Fermi energy.

All results reported here use the short-wavelength cutoff approximation introduced by Samalam and Serene~\cite{sam78}, which we have discussed and justified in Ref.~\cite{lin21}. In the following formulas the parameter $x_{\mathrm{c}}$ is the dimensionless cut-off of the total Cooper pair momentum $q$, measured in the unit of $1/\xi_0$, which we expect to be of order unity. 

The result for the correction to the Fermi velocity, relative to the Fermi liquid result, is
\begin{multline}\label{vf-correction}
\frac{\delta v_{\mathrm{f}}}{v_{\mathrm{f}}}=
-\frac{4.7174}{2^{10}}\frac{1}{N(0)T_{\mathrm{c}}\xi_{0}^{3}}\times
\Big[\frac{\arctan\frac{x_{\mathrm{c}}}{\sqrt{\vartheta}}}{\sqrt{\vartheta}^{3}}+
\frac{x_{\mathrm{c}}^{3}-\vartheta x_{\mathrm{c}}}{\vartheta x_{\mathrm{c}}^{4}+2\vartheta^{2}x_{\mathrm{c}}^{2}+\vartheta^{3}} \\
  -\frac{\pi^{2}}{7\zeta(3)}\Big(\frac{3\arctan\frac{x_{\mathrm{c}}}{\sqrt{\vartheta}}}{\sqrt{\vartheta}}-
  \frac{5x_{\mathrm{c}}^{3}+3\vartheta x_{\mathrm{c}}}{x_{\mathrm{c}}^{4}+2\vartheta x_{\mathrm{c}}^{2}+\vartheta^{2}}\Big)\Big]\\
  -5.975\times10^{-2}\times\frac{1}{N(0)T\xi_{0}^{3}}[x_{\mathrm{c}}-\sqrt{\vartheta}\atanfactor].
\end{multline}
The coefficient of the time derivative is written as $1+\delta$, with
\begin{multline}\label{dt-correction}
\delta = 
\frac{4.7174}{2^{10}}\frac{1}{N(0)T_{\mathrm{c}}\xi_{0}^{3}}\times\Big[\frac{\arctan\frac{x_{\mathrm{c}}}{\sqrt{\vartheta}}}{\sqrt{\vartheta}^{3}}+\frac{x_{\mathrm{c}}^{3}-\vartheta x_{\mathrm{c}}}{\vartheta x_{\mathrm{c}}^{4}+2\vartheta^{2}x_{\mathrm{c}}^{2}+\vartheta^{3}} \\
-\frac{\pi^{2}}{7\zeta(3)}\Big(\frac{3\arctan\frac{x_{\mathrm{c}}}{\sqrt{\vartheta}}}{\sqrt{\vartheta}}-\frac{5x_{\mathrm{c}}^{3}+3\vartheta x_{\mathrm{c}}}{x_{\mathrm{c}}^{4}+2\vartheta x_{\mathrm{c}}^{2}+\vartheta^{2}}\Big)\Big].
\end{multline}
Similarly, the corrections to the Landau interaction parameters, $F^s_{0,1}$, reduce to
\begin{equation}\label{F0}
\begin{split}
\delta F^s_0 =
-& 4.617\times10^{-3}\frac{1}{N(0)\xi_{0}^{3}T_{\mathrm{c}}}
\Big[\frac{\atanfactor}{\sqrt{\vartheta}}-\frac{x_{\mathrm{c}}}{x_{\mathrm{c}}^{2}+\vartheta}\Big]\\
+& 3.136\times10^{-2}\frac{1}{N(0)T_{\mathrm{c}}\xi_{0}^{3}}
\Big[x_{\mathrm{c}}-\frac{3}{2}\sqrt{\vartheta}\atanfactor
+ \frac{\vartheta x_{\mathrm{c}}}{2(x_{\mathrm{c}}^{2}+\vartheta)}\Big] \\
+& 5.975\times10^{-2}\times\frac{1}{N(0)\xi_{0}^{3}T_{\mathrm{c}}}
\Big[x_{c}-\sqrt{\vartheta}\atanfactor\Big]
\,,
\end{split}
\end{equation}
and
\begin{equation}\label{F1}
\begin{split}
\delta F^s_1 =
&1.211\times10^{-2}\frac{1}{N(0)T_{\mathrm{c}}\xi_{0}^{3}}
\Big[x_{\mathrm{c}}-\frac{3}{2}\sqrt{\vartheta}\atanfactor
+ \frac{\vartheta x_{\mathrm{c}}}{2(x_{\mathrm{c}}^{2}+\vartheta)}\Big] \\
-& 1.792\times10^{-1}\times\frac{1}{N(0)\xi_{0}^{3}T_{\mathrm{c}}}
\Big[x_{c}-\sqrt{\vartheta}\atanfactor\Big].
\end{split}
\end{equation}
These corrections can be used to calculate fluctuation corrections to other thermodynamic and transport properties. In particular, virtual emission and absorption of Cooper pairs results in a correction to the zero sound velocity.

\section{Corrections to the Zero Sound Velocity}\label{sec-zs}

Zero sound is a collective excitation, an eigenmode of the collisionless kinetic equation corresponding to coherent particle-hole excitations near the Fermi surface. Zero sound propagates at low temperatures in the ``collisionless limit'', $\omega\tau(T)\gg 1$, weakly damped by collisions with thermal excitations. 
In this limit the restoring force for the deformation of the Fermi surface is provided by the Landau interaction, particularly the interaction energy associated with a dilatation of the Fermi surface, i.e. $F^s_0$, and that associated with the dipole mode, $F^s_1$. The quadrupolar distortion of the Fermi surface, and thus $F^s_2$, also contributes to the zero sound mode.
Zero sound was predicted by Landau~\cite{lan57}, and played an important role in establishing liquid \he{} as a Landau Fermi liquid, with the experimental observation of zero sound, including the predictions for the velocity and attenuation, by Abel, Anderson, and Wheatley~\cite{abe66}.

Theoretically, it can be shown that to leading order in $1/F^s_0 \ll 1$ the zero sound velocity $c_0$ is given by~\cite{baym91}

\begin{equation}\label{zero-sound_velocity}
\frac{c_0^2}{c_1^2} - 1 = \frac{4}{5}\frac{1+\frac{1}{5}F^s_2}{1+F^s_0},
\end{equation} 
where $c_1/v_{\mathrm{f}}=\sqrt{\frac{1}{3}(1+F_0^s)(1+F_1^s/3)}$ is the 
hydrodynamic sound velocity~\cite{baym91}.

The above relation for $c_0$ also depends on the Landau parameter $F^s_2$, which is of order $|F^s_2|\sim\cO(1)$, but its precise value is not well established~\cite{dobbs00,har00}.

Based on the magnitude of the corrections given in the previous section, combined with Eq.~\eqref{zero-sound_velocity}, we observe that the Landau parameter corrections to the zero sound velocity are much smaller than the corrections for the time-derivative coefficient and Fermi velocity. We neglect $F^s_2$ in the above formula, and obtain
\begin{equation}
\frac{c_0^2}{v_{\mathrm{f}}^2} = \frac{1}{3}\left(1+\frac{1}{3}F_1^s\right)\left(1+F_0^s + \frac{4}{5}\right).
\end{equation}
Using the results for the corrections to the Fermi velocity and dispersion relation given in the previous section, the zero sound velocity in the presence of pair fluctuations is given by
\begin{equation}
c_0^{\text{fluc}} = 
\frac{v_{\mathrm{f}} +\delta v_{\mathrm{f}}}{1+\delta}
\sqrt{\frac{1}{3}\left(1+\frac{1}{3}F_1^s\right)\left(1+F_0^s + \frac{4}{5}\right)}
\,.
\end{equation}

\begin{figure}[tbp]
\centerline{\includegraphics[width=1.0\textwidth]{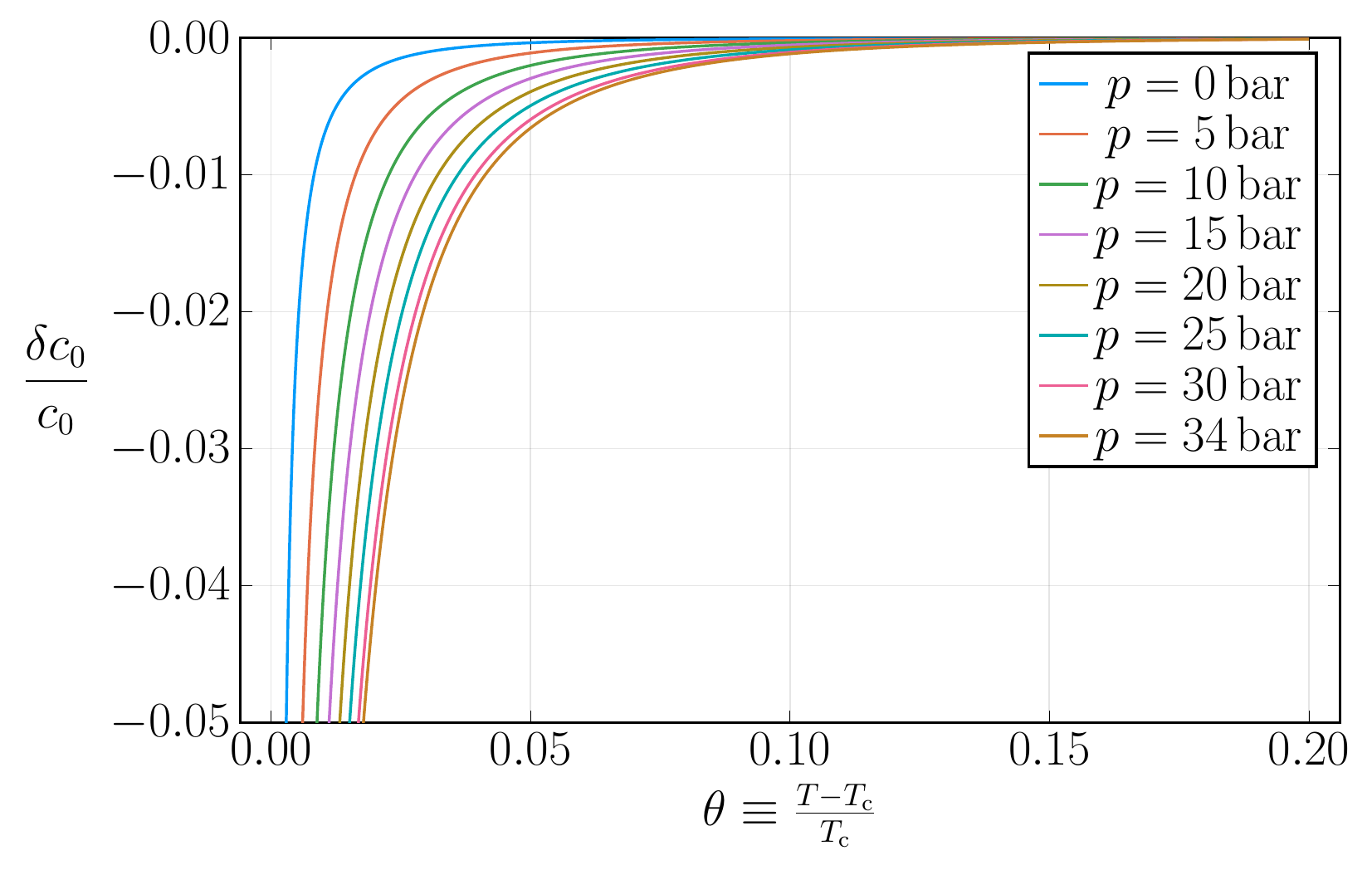}}
\caption{The deviation of the zero sound velocity relative to its Fermi liquid value, $\delta c_0/c_0$, at different pressures. The curves are calculated with a fixed cut-off $x_{\mathrm{c}} = 0.236$.}
\label{fig-zero-sound-speed}
\end{figure}

In Fig.~\ref{fig-zero-sound-speed} we show the result for the correction to the zero sound velocity $\delta c_0\equiv c^{\text{fluc}}_0 - c_0$, where $c_0$ is the zero sound velocity in the absence of pairing fluctuations. We use a cut-off of $x_{\mathrm{c}}=0.236$ obtained from the fit of our approximate theoretical result for the fluctuation correction to the experimental measurements of the excess zero sound attenuation reported by Paulson and Wheatley~\cite{pau78c}, as discussed in Ref.~\cite{lin21}.
Note the rapid suppression of the sound velocity for $T\rightarrow T_{\mathrm{c}}$, which becomes more pronounced with increasing pressure. The relative change in velocity is of order $10^{-2}$, arising 
from the corrections to Fermi velocity and the dispersion correction, $\delta$.
The corrections to $c_0$, from $\delta F_l^s$, are of the order $10^{-4}$, and are insignificant for the zero sound velocity.

This result should be observable with precision measurements of the sound velocity. Lee et~al.~\cite{lee96} do report measurements of the zero-sound velocity for temperatures close to $T_{\mathrm{c}}$. Their result indeed shows a velocity drop near $T_{\mathrm{c}}$, but the magnitude of the drop they report is approximately $30\,\mbox{ppm}$, several orders of magnitude smaller than our theoretical prediction shown in Fig.~\ref{fig-zero-sound-speed}.

We are not able to resolve the discrepancy based on the limited data, but we speculate on a possible explanation.
Our analysis of the correction to the sound velocity is based on a thermal population of Cooper pair fluctuations. However, the frequency of zero sound in the experiment of Ref.~\cite{lee96} is $\omega/2\pi=113\,\mbox{MHz}$, which is the same magnitude as the rate of Cooper pair formation, i.e. $\hbar\omega/2\pi\kb T_{\mathrm{c}}\approx 0.86$. This suggests that zero sound likely suppresses the formation of pair-fluctuations, and as a result the fluctuation correction to the sound velocity. To test this hypothesis requires an extension of the theory to calculate the nonequilibrium population of Cooper pair fluctuations in the presence of high frequency zero sound. The bottom line is that additional research is necessary.

\section{Summary}

We have developed a theory for the effects of Cooper pair fluctuations on Fermi systems starting from nonequilibrium Keldysh field theory. Here we report our results for the leading corrections to the collisionless Boltzmann-Landau kinetic theory for liquid \he{}, including the Fermi velocity, Landau parameters and zero sound velocity.
There are very few studies of fluctuation effects in the Fermi liquid region just above $T_{\mathrm{c}}$. We hope our theoretical predictions will motivate new research on the fluctuation region near $T_{\mathrm{c}}$.

\paragraph{Acknowledgments}
This research was supported by the National Science Foundation Grant DMR-1508730.

\paragraph{Data} supporting this report is available on request.

\medskip
\noindent $^*$ Corresponding author: sauls@northwestern.edu.

\section*{Appendix: Expressions for $\Gamma_Q$ and $\Sigma^{11}$}
We give the expressions for the fluctuation vertex $\Gamma_Q$ and the related self-energy $\Sigma^{11}$, which is needed for the correction to quasiparticle energy.
The derivation is given in details in \cite{lin21}.
The effective vertex $\Gamma_Q$ is defined by the Bethe-Salpeter equation in the particle-particle channel,
\begin{multline}
\Gamma_{\alpha\beta, \gamma\delta}^{ab}(p,p';Q) =
iV_{\alpha\beta, \gamma\delta}(p,p')\check{\tau}_3^{ab} \\
+ \sum iV_{\alpha\beta, \alpha'\beta'}(p,p'')\check{\tau}_3^{ac}
G^{cd}(p'')G^{cd}(Q-p'')
\Gamma_{\alpha'\beta',\gamma\delta}^{db}(p'', p';Q),
\end{multline}
where $V$ is the pairing interaction, and the upper indices are for the forward and backward branches in Keldysh formalism.
The variable $Q = (\V{q}, \omega)$ represents the momentum and energy of the pair fluctuations.
The vertex can be decomposed into orbital and spin parts,
\begin{equation}
\Gamma_{\alpha\beta, \gamma\delta}^{ab}(p,p';Q) = \Gamma^{ab}(p,p';Q)
\times 
\frac{1}{2}\V{g}_{\alpha\beta}\cdot\V{g}^\dagger_{\gamma\delta} \,,
\end{equation}
where $\V{g} \equiv i\boldsymbol{\sigma}\sigma_y$ is the triplet of spin-symmetric matrices for spin-triplet pairing.
Define
\begin{align}
\label{GG12}
(GG)^{12} &= -2\pi \int_\mathbf{p}\delta(\omega - \eps_\mathbf{p} - \eps_\mathbf{q-p})
n_{\mathbf{p}}n_{\V{q}-\V{p}}
\hat{p}_i \hat{p}_k,
\\
\label{GG21}
(GG)^{21} &= -2\pi \int_\mathbf{p} \delta(\omega - \eps_\mathbf{p} - \eps_\mathbf{q-p})
\bar{n}_{\mathbf{p}}\bar{n}_{\V{q}-\V{p}}
\hat{p}_i \hat{p}_k,
\end{align}
and
\begin{equation}\label{LGL}
L_{ik} \equiv \frac{1}{3iV} 
- i\int_\mathbf{p} 
\frac{n_{\mathbf{p}}+n_{\V{q}-\V{p}}-1}{\omega - \eps_\mathbf{p}-\eps_\mathbf{q-p}+i0}
\hat{p}_i \hat{p}_k \,,
\end{equation}
where $n_{\V{p}}$ is the quasiparticle distribution function and $\bar{n}_{\V{p}}\equiv 1- n_{\V{p}}$.
Using the projections $P_{ij}^{\parallel}=\hat{q}_i\hat{q}_j$ and $P_{ij}^{\perp}=\delta_{ij}-\hat{q}_i\hat{q}_j$ along the direction $\hat{\V{q}}$ of the pair fluctuations, we can decompose the above quantities into $\parallel$ and $\perp$ components: $(GG)^{ac}_{ik}=\sum_{\lambda=\parallel,\perp} (GG)^{ac}_\lambda P_{ik}^{\lambda}$ and $L_{ik}=\sum_{\lambda} L^{\mathrm{GL}}_{\lambda} P^{\lambda}_{ik}$.
The orbital part of the effective vertex describing pair fluctuations is given by
\begin{equation}
\Gamma_\lambda 
=\frac{1}{|L^{\mathrm{GL}}_\lambda|^{2} }
\begin{bmatrix}
-L^\mathrm{GL}_\lambda - (GG)^{21}_\lambda & (GG)^{12}_\lambda \\
(GG)^{21}_\lambda & L_\lambda^\mathrm{GL} - (GG)^{12}_\lambda
\end{bmatrix}
\end{equation}
for $\lambda=\parallel,\perp$.

In equilibrium, the $L$ function is given by 
\begin{equation}\label{singular}
L^{\mathrm{GL}}_\lambda = i\frac{N(0)}{3}\Big(\vartheta + \xi_\lambda^2 q^2 - i\frac{\pi}{8}\frac{\omega}{T}\Big),
\end{equation}
with the reduced temperature $\vartheta\equiv T/T_{\mathrm{c}}-1$ and the coherence length scales, $\xi_\parallel^2 = \frac{9}{5}\xi_0^2$ and $\xi_\perp^2 = \frac{3}{5}\xi_0^2$, where $\xi_0^2 = \frac{7\zeta(3)}{48\pi^2}\frac{v_\mathrm{F}^2}{T_\mathrm{c}^2}$.
The singularity at $Q=0$ as $T\to T_{\mathrm{c}}$ signifies the onset of the phase transition.

The (1,1) component of the self-energy given by the diagram Fig.~\ref{fmf-Self_Energies}(d) is
\begin{equation}
\Sigma^{11}(p)=\sum_Q \Gamma^{11}_{\alpha\beta,\gamma\delta}(p,p;Q)G^{11}_{\beta\delta}(Q-p).
\end{equation}
This is the only relevant component for this article (see Equation~\eqref{linear-BLE}).
More details are given in~\cite{lin21}.


\end{document}